%% file: samplepaper.tex
\documentclass[runningheads]{llncs}
\usepackage{amsmath}
\usepackage{algorithm}     
\usepackage{algpseudocode}
\usepackage{enumitem}
\usepackage{float}
\usepackage{booktabs} 
\usepackage{multirow}

\usepackage{booktabs} % 提供更好看的横线
\usepackage{colortbl} % 用于背景颜色
\usepackage{xcolor}   % 定义颜色

\usepackage{adjustbox}
\usepackage{graphicx}
\usepackage[table,xcdraw]{xcolor}
\usepackage{amsmath}
\usepackage{caption}
\usepackage{makecell}
\usepackage{enumitem}
\usepackage{booktabs}

\usepackage{tabularx}
\usepackage{array}

\usepackage[T1]{fontenc}
\usepackage{graphicx}
\begin{document}
\title{Curr-RLCER:Curriculum Reinforcement Learning For Coherence Explainable Recommendation}
% \author{Anonymous Submission}
% \institute{}

\author{Xiangchen Pan\inst{1} \and Wei Wei\inst{1}\thanks{Corresponding author.} }
\authorrunning{X. Pan et al.}
\institute{Huazhong University of Science and Technology, Wuhan, China \\
\email{pxcstart666@gmail.com} \email{weiw@hust.edu.cn}\\
}

\maketitle              % typeset the header of the contribution
\begin{abstract}
Explainable recommendation systems (RSs) are designed to explicitly uncover the rationale of each recommendation, thereby enhancing the transparency and credibility of RSs. Previous methods often jointly predicted ratings and generated explanations, but overlooked the incoherence of such two objectives. To address this issue, we propose Curr-RLCER, a reinforcement learning framework for explanation coherent recommendation with dynamic rating alignment. It employs curriculum learning, transitioning from basic predictions (i.e., click through rating-CTR, selection-based rating) to open-ended recommendation explanation generation. In particular, the rewards of each stage are designed for progressively enhancing the stability of RSs. Furthermore, a coherence-driven reward mechanism is also proposed to enforce the coherence between generated explanations and predicted ratings, supported by a specifically designed evaluation scheme.  The extensive experimental results on three explainable recommendation datasets indicate that the proposed framework is effective. Codes and datasets are available at https://github.com/pxcstart/Curr-RLCER.

\keywords{Explainable Recommendation \and Reinforcement Learning \and Curriculum Learning.}
\end{abstract}

\input{section/introduction}

\input{section/methodology}
\input{section/experiment}

\input{section/related}

\input{section/conclusion}


\bibliographystyle{splncs04}
\bibliography{reference}





% LNCS Homepage, \url{http://www.springer.com/lncs}, last accessed 2023/10/25
% \end{thebibliography}
\end{document}

%% file: section/introduction.tex
\section{Introduction}
Recommendation systems have always been an important part of data mining and information retrieval, recommending items that match user preferences by learning and predicting their preferences. In recent years, in order to alleviate the credence matters caused by the lack of transparency and explainability in recommendation systems, explainable recommendation has received increasing attention. By inferring the fundamental reasons behind the historical interactions between users and items, recommendation models can provide explainable suggestions and insights into the decision-making process, enabling more personalized and transparent recommendations.
% 推荐系统一直以来是数据挖掘和信息检索中重要的一环，通过学习和预测用户的偏好，向用户推荐符合用户偏好的物品。近年来，为了缓解推荐系统因缺乏透明度和可解释性而导致的信任问题，可解释推荐受到了越来越多的关注，通过对用户和商品的历史交互背后的根本原因的推理，推荐模型能够提供具有可解释性的建议和决策过程的见解，从而进行更加个性化、更加透明化的推荐。

At present, most of the work on explainable recommendations is carried out in a multitasking manner, where the system generates ratings and provides corresponding explanations. For example, Att2Seq~\cite{dong2017learning} and NRT~\cite{li2017neural} use attention mechanisms and recursive neural networks (RNNs) to generate text explanations. Recent advances have further explored the utilization of Transformers in text generation, providing valuable insights for recommendation results. Although these methods can generate richer and smoother text explanations compared to early methods based on predefined sentence templates~\cite{zhang2014explicit,li2021caesar}, they focus on improving the quality of text generation and overlook the inconsistency between predicted ratings and generated explanations, which will greatly weaken the credibility of the system.
% 目前关于可解释推荐的工作大多采用多任务的方式开展，系统会产生评级并提供相应的解释。例如，Att2Seq 和 NRT 采用注意力机制和递归神经网络（RNN）来生成文本解释。最近的进展进一步探索了Transformer在文本生成中的利用，为推荐结果提供了有价值的见解...尽管相比于早期基于预定义句子模板的方法可以产生更丰富、更流畅的文本解释，但这些方法关注如何提高文本生成的质量，却忽视了在预测评级和生成的解释之间存在不一致性，这将大大削弱系统的可信度。

 In order to improve the consistency between predicted ratings and generated explanations, CER~\cite{raczynski2023problem} adds a consistency alignment strategy on the basis of the PETER framework. By enhancing the consistency between predicted ratings and explanation-based ratings generated by max pooling, the training model can generate more personalized and consistent explanations. However, these methods rely on tools to extract sentiment quadruples from user reviews, and the quality of the data can seriously affect the training effectiveness of the model, thereby affecting the quality of the generated explanations. Moreover, the existing coherence alignment and evaluation strategies have ignored the situation where both predicted ratings and generated explanations are distorted. 
% 为了提高预测评级和生成解释之间的一致性，CER在PETER框架的基础上增加了一致性对齐策略，通过对预测评分与根据解释进行最大池化生成的打分进行一致性增强，训练模型能够生成更加个性化和一致的解释。然而这些方法都依赖于借助工具从用户评论中提取情感四元组，数据的质量会严重影响模型的训练效果，进而影响生成的解释质量。并且目前已有的一致性对齐和评估策略均忽视了预测评分和生成解释同时失真的情况

% In order to improve the consistency between ratings and explanations, CER has added a consistency alignment strategy based on the PETER framework. By enhancing the consistency between predicted scores and scores generated by max pooling based on explanations, the training model can generate more personalized and consistent explanations; CIER has made improvements in the backbone model and training techniques, using LLM to predict ratings and generate explanations through two-stage SFT fine-tuning, and pointing out that the training method of multi-task learning only sharing hidden representation layers but independent output layers is the reason for inconsistent explanations. However, when using SFT fine-tuning for explainable recommendations, the training performance of the model heavily depends on the quality of annotated data. For LLMs with small parameter quantities, this coherence enhancement lacks generalization and has a significant Out-of-Domain (OOD) problem.
% 为了提高评级和解释之间的一致性，CER在PETER框架的基础上增加了一致性对齐策略，通过对预测评分与根据解释进行最大池化生成的打分进行一致性增强，训练模型能够生成更加个性化和一致的解释；CIER则在主干模型和训练技巧上进行了改进，使用LLM通过二阶段sft微调的方式来预测评级并生成解释，并指出多任务学习只共享隐藏表示层但独立输出层的训练方式是导致不一致解释的原因。然而通过sft微调进行可解释推荐，模型的训练效果会严重依赖注释数据的质量，并且对于参数量较小的大模型而言，这种一致性增强缺乏泛化性，存在较为明显的OOD（Out-of-Domain）问题。

Recently, the successful case of Deepseek R1-Zero~\cite{guo2025deepseek} enhancing LLM inference through Group Relative Policy Optimization (GRPO) demonstrates that by designing appropriate reward models, LLMs can still acquire deep thinking abilities through continuous self validation and have good inference performance without performing SFT fine-tuning but by directly training the basic model with reinforcement learning, thereby reducing dependence on high-quality supervised data. Given the effectiveness of GRPO in inference tasks, we attempt to use reinforcement learning based LLM training to solve the problem of inconsistency between rating and explanation in explainable recommendation.
% 近来，Deepseek R1-Zero通过组相对策略优化（GRPO）增强大模型推理的成功案例说明，通过设计合适的奖励模型，大模型可以在不断地自我验证的过程中获得深度思考的能力，并且在不进行sft微调，直接对基础模型进行强化学习训练的情况下，模型依然可以具备良好的推理性能，从而减轻对高质量监督数据的依赖。鉴于GRPO在推理任务中的有效性，我们尝试采用基于强化学习的大模型训练来缓解评分和解释不一致的问题。

Considering the disaster of the forgetting problem of LLMs in multi-task learning, relevant studies~\cite{deng2025boosting} have shown that curriculum learning can effectively alleviate loss oscillations during training. Specifically, we let the LLM first learn the CTR prediction task, which is to determine whether the user likes the item. The goal is for the LLM to adapt to recommendation data and learn how to make recommendations; In the second stage, we designed a rating prediction task, allowing the LLM to play the role of a consumer and rate for items, further enhancing the model's rating prediction ability; In the third stage, we designed an explanation generation task to enable the LLM to generate explanations for the recommendation based on user and item profile information, as well as reviews, in order to enhance the explainability of the model. In three stages, we transitioned from simple binary classification to multi classification tasks and from discriminative tasks to generative tasks. We designed corresponding reward functions for each stage, gradually increasing the complexity of the task from easy to difficult.
% 考虑到大模型在多任务学习时的灾难遗忘问题，相关研究表明，课程学习可以有效的减轻训练中的损失震荡。具体来说，我们让大模型首先学习CTR预测任务，即判断用户对商品是否喜欢，目标在于大模型适应推荐数据，学会做推荐；在第二阶段，我们设计了评分预测任务，让大模型扮演消费者，对商品进行打分，进一步增强模型的评分预测能力；在第三阶段，我们设计了解释生成任务，让大模型根据用户和商品配置信息以及评论，生成对该推荐的解释，提升模型的可解释能力。三个阶段从简单的二分类过渡到多分类任务，从判别式任务过渡到生成式任务，每个阶段我们设计了相应的奖励函数，由易到难的逐步增加任务的复杂性。

To more effectively train and evaluate the coherence between model-generated explanations and their corresponding ratings, we introduce a coherence reward mechanism and a coherence evaluation framework. In the explainable training phase, we will utilize a pre-trained sentiment classifier to rate explanations and compare explanation-based ratings with the ground truth ratings for coherence as part of the reward mechanism. During the inference evaluation phase, we employ a multi-faceted evaluation approach to comprehensively assess the coherence between explanations and ratings. Specifically, we utilize (i) a sentiment classifier to provide automated sentiment assessment, (ii) a large language model (e.g., GPT) for contextual judgment, and (iii) human evaluators for qualitative analysis. The explanation-derived ratings obtained from these sources are compared with both the predicted ratings and the ground-truth ratings to assess consistency. 

% In order to better train and evaluate the coherence between explanations and ratings, we propose a consistency reward mechanism and consistency evaluation scheme. In the explainable training phase, we will utilize a pre-trained sentiment classifier to rate explanations and compare explanation-based ratings with the ground truth ratings for coherence as part of the reward mechanism; In the inference evaluation stage, we use sentiment classifier, GPT, and manual evaluation to rate explanation and compare explanation-based rating with both predicted rating and ground truth rating for comprehensive coherence evaluation.
% 为了更好的训练和评估解释与评分之间的一致性， 我们提出了一套一致性奖励机制和一致性评估方案。在可解释训练阶段，我们会对解释进行情感打分，与真实得分进行一致性比较，作为奖励机制的一部分；在推理评估阶段，我们使用该情感分类模型、GPT3.5和人工评估，对解释进行情感打分，已有的评估方法着重于考虑解释与评分的情感连贯性，却忽视了二者同时失真的情况，我们将基于解释的评分与预测评分、真实评分一起进行综合考量，提出了一种综合一致性评估方案。

The following is a summary of the contributions of this work:

\begin{itemize}[leftmargin=*, noitemsep] 
    \item We propose a reinforcement learning based LLM training framework, Curr-RLCER, to address the issue of inconsistent ratings and explanations in explainable recommendations. 
    \item We have customized a coherence reward mechanism and coherence evaluation scheme, taking into account the distortion of predicted ratings and explanation-based ratings in the evaluation process.
    \item Using three benchmark datasets for experiments, our method significantly improves the coherence of explanation compared to other state-of-the-art methods, without reducing other metrics for recommendation predictions.
\end{itemize}
% 下面是关于本工作的贡献总结：
% （1）针对可解释推荐中评分和解释不一致的问题，我们提出了一种基于强化学习的大模型训练框架Curr-RLCER，该框架采用课程学习的方式进行多任务学习，缓解了传统方法严重依赖高质量注释数据和泛化性差的问题。
% （2）我们自定义了一种一致性奖励机制和一致性评估方案，在评估方面，考虑了预测评级和解释评级同事失真的情况。
% （3）使用三个基准数据集进行实验，我们的方法与其他最先进的方法相比，能够显著提高解释的连贯性，而不会减少推荐预测的其他度量。

%% file: section/methodology.tex
\section{Preliminaries}
Given a set of users $U$, a set of items $I$, and a list of reviews $T_{u, i}$, the goal of this work is to utilize user profile $p_{u, m}$ of user $u_m$ and the item profile $p_{i, n}$ of item $i_n$, as well as review data $t_{u_m,i_n}$, to predict ratings $r_{u_m,i_n}$ and generate explanations $E_{u_m, i_n}$. Let's further assume that the rating is a score $r_{u_m, i_n} 
\in \{1,2,3,4,5\}$ to measure the consistency between the item attributes and user preferences. When evaluating the explanations generated by the model, we not only consider the text quality of the explanation and the stability of the generation but also consider whether it is consistent with the predicted score $r_{u_m,i_n}$.
% 给定用户集$U$和商品集$I$ 以及评论列表$T_{u,i}$,本工作中的目标是利用用户$u_m$的配置信息$p_{u_m}$和商品$i_n$的配置信息$p_{i_n}$以及评论数据$t_{u_m,i_n}$来预测评分$r_{u_m,i_n}$并生成解释$E_{u_m,i_n}$。让我们进一步假设评分是一个分数$r_{u_m,i_n} \in \{1,2,3,4,5\}$来衡量商品和用户偏好的一致性。在评估模型生成的解释时，我们不仅考虑解释的文本质量和生成的稳定性，还考虑了它是否与预测评分$r_{u_m,i_n}$是否具有一致性。

\begin{figure*}[t]
    \centering
    \includegraphics[width=\linewidth]{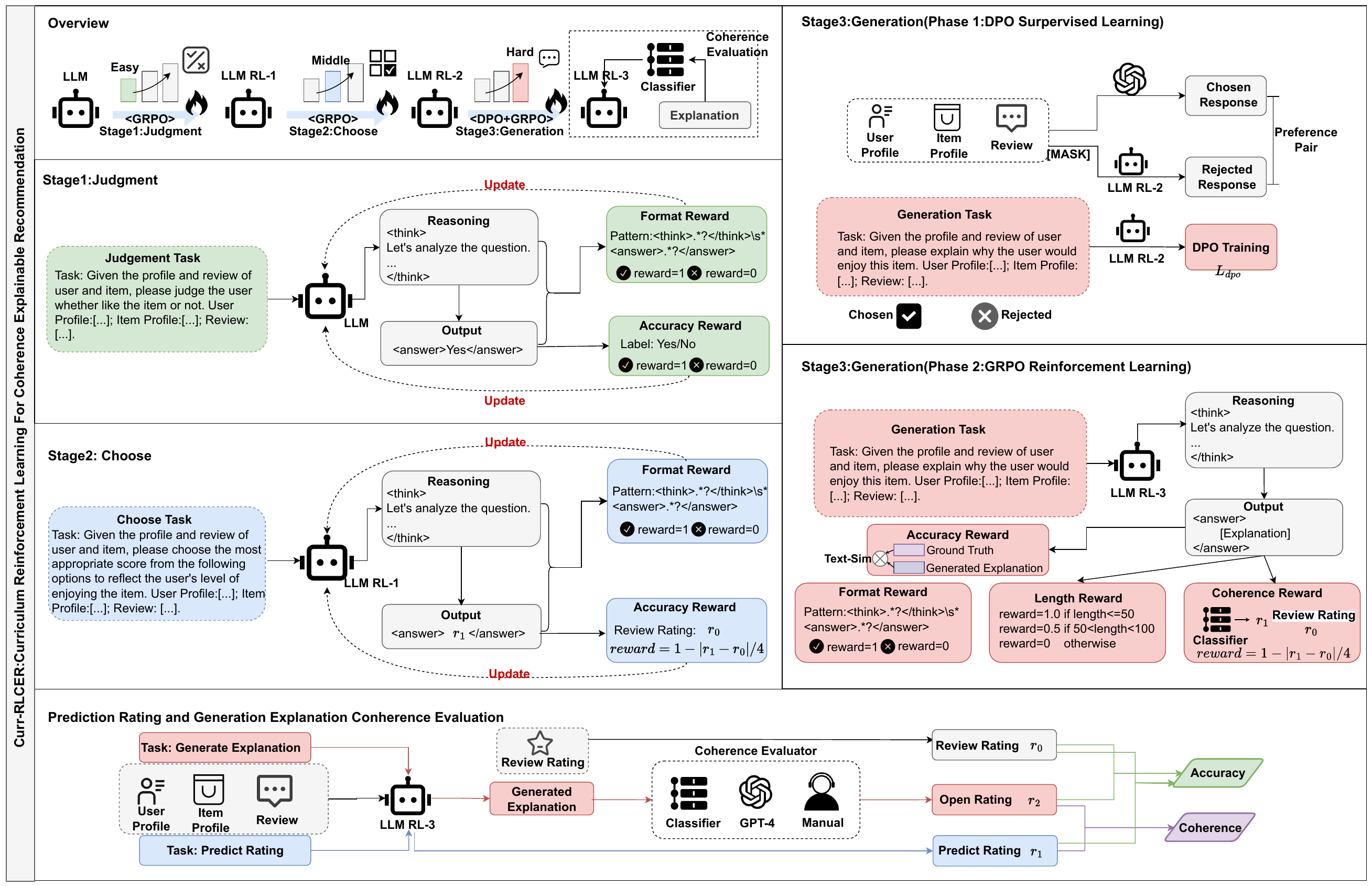}
    \caption{The overview framework of Curr-RLCER.}   
    \label{fig1}
\end{figure*}

\section{Methodology}
In this section, we will briefly introduce and provide a detailed explanation of the Curr-RLCER framework, an explanation-rating coherence enhanced recommendation framework based on three-stage curriculum reinforcement learning. The overall framework is shown in Fig \ref{fig1}. In three stages, we transitioned from simple binary classification to multi classification tasks and from discriminative tasks to generative tasks. We designed corresponding reward functions for each stage, gradually increasing the complexity of the task from easy to difficult. We will discuss the DPO and GRPO algorithms used in Curr-RLCER in Section 3.1 and introduce the specific implementation of the reward mechanism for each stage in Section 3.2. In addition, we have designed a comprehensive coherence assessment scheme, as detailed in Section 3.3.
% 在这一节中，我们将简要介绍下详细说明Curr-RLCER框架，一个基于三阶段课程强化学习的解释-评级一致性增强推荐框架。整体框架如图*所示。三个阶段从简单的二分类过渡到多分类任务，从判别式任务过渡到生成式任务，每个阶段我们设计了相应的奖励函数，由易到难的逐步增加任务的复杂性。我们将在Sec 4.1中讲述框架中使用到的DPO和GRPO算法，并在Sec 4.2中介绍每个阶段奖励机制的具体实现。此外，我们还设计了一种综合一致性评估方案，具体可见Sec 4.3。

% Curr-RLCER包含三个训练阶段，第一阶段通过判断形式的CTR预测任务来提升模型的基本推荐能力，第二阶段通过选择形式的评级预测任务提升模型的评级预测能力；第三阶段首先通过dpo训练使模型何为高质量推荐，然后再通过基于一致性奖励机制的grpo训练提升评级-解释的一致性。框架图的最下方为本方法中综合一致性评估方案的流程示意图。

\subsection{Reinforcement Learning for LLM}
\subsubsection{DPO}
Reinforcement learning based on human feedback (RLHF) is a key approach for aligning machine learning models with human preferences. Among these techniques, the Direct Preference Optimization (DPO) algorithms have gained widespread attention due to their excellent alignment performance and simple and efficient design. We use the strategy $\pi$ to represent an LLM, which, for a given input $x$, generates a response $y$ from the distribution $\pi(\cdot|)$. Assuming the initial strategy is $\pi_0$ and the preference dataset is $D={(x, y_w, y_l)}$, where for each input x, the priority of response $y_w$ is higher than the response $y_l$. The objective function of DPO is based on KL divergence to measure the difference between the new strategy and the reference strategy while considering preference rewards:
% 来自人类反馈的强化学习（RLHF）是使机器学习模型与人类偏好保持一致的关键方法，在这些技术中，直接偏好优化(DPO)算法凭借其良好的对齐性能和简单高效的设计广受关注。我们用策略$\pi$表示一个LLM，对于给定的输入x，从分布$\pi(.|x)$中生成一个回复y。假设初始策略为$\pi_0$，偏好数据集为$D={(x,y_w,y_l)}$，其中对于每个输入x，回复$y_w$的优先级高于回复$y_l$。DPO 的目标函数基于 KL 散度来衡量新策略与参考策略之间的差异，同时考虑偏好奖励：
\begin{multline}
L_{DPO} = -E_{(x,y,y_w,y_l) \sim D} \\
\Bigg[ \log \sigma \bigg( \beta \log \frac{\pi_\theta(y_w|x,v)}{\pi_0(y_w|x,v)} - \beta \log \frac{\pi_\theta(y_l|x,v)}{\pi_0(y_l|x,v)} \bigg) \Bigg]
\end{multline}

\subsubsection{GRPO}
Reinforcement learning based on human feedback heavily relies on the evaluation of critic models. Recently, Deepseek R1-Zero introduced the GRPO framework, which eliminates dependence on other critical models. Specifically, GRPO considers the relative performance of responses rather than absolute reward values. For a given input query $q$, GRPO generates $N$ different responses $\{o_1, o_2,..., o_N\}$ from the current policy $\pi_{\theta_{old}}$, and evaluates through group-wise comparisons:
% 基于人类反馈的强化学习严重依赖critic model的评估，最近，Deepseek R1-Zero引入了GRPO框架，消除了对其他critic model的依赖，具体来说，GRPO考虑的是响应的相对表现，而不是绝对奖励值。对于给定的输入查询q，GRPO从当前策略$\pi_\theta_{old}$中生成N个不同的响应${o_1,o_2,...,o_N}$,并通过分组比较进行评估：
\begin{equation}
\begin{aligned}
A_i = \frac{r_i-mean({r_1,...,r_N})}{std({r_1,...,r_N})}
\end{aligned}
\end{equation}
where $A_i$ represents the normalized relative quality of the i-th response within its group.
% 其中$A_i$表示组内第i个响应的标准化相对质量。

\subsection{Curriculum Reinforcement Learning}
In order to alleviate the instability of reinforcement learning, we integrated curriculum learning with GRPO, trained the LLMs in stages, and set a difficulty perception reward mechanism consistent with progress for each stage of the task. Specifically, reinforcement learning training consists of three stages: binary classification, multiple choice, and open-ended response. In terms of task format, the task in the binary classification stage is the CTR prediction task, which determines whether the user likes the item; the task in the multiple-choice stage is the rating prediction task, which involves scoring items to predict users' level of enjoyment; and the task in the open-ended response stage is to generate explanations. 
% 为了缓解强化学习不稳定的问题，我们采用课程学习的方式与GRPO进行集成，对大模型分阶段的进行训练，并为每个阶段的任务设置了与进展相一致的难度感知奖励机制。具体来说，强化学习训练由三个阶段组成：二元分类、多项选择和开放式问答。任务形式上，二元分类阶段的任务为CTR预测任务，即判断用户是否喜欢商品；多项选择阶段的任务为评分预测任务，即对商品进行打分；开放式问答阶段的任务为解释生成任务，即生成推荐的解释。

% Table \ref{table1} shows the prompts used for model training at various stages. In the process of data construction, user/item profile and ground truth explanations are generated through GPT-3.5-Turbo, We referred to the work of XRec~\cite{ma2024xrec} for the prompt constructing. Below is a detailed introduction to the reward design and training details for each stage.
% 下面详细介绍每个阶段的奖励设计与训练细节。

\subsubsection{Stage1: Judgment for CTR Recommendation}
At this stage, we use user and item profile information, as well as reviews, as input data. The task instruction is "judge the user whether they like the item or not." In addition, we explicitly require the model to respond with "yes" or "no" in the instruction. We design corresponding reward functions based on the correctness of the feedback results from the model:
% 在该阶段，我们将用户和商品的配置信息以及评论作为输入数据，任务指令为"judge the user whether like the item or not"，此外，我们在指令中还明确要求了模型必须以“yes”或“no”的方式进行响应，详细的提示词信息可见附录A.1。我们根据模型反馈的结果正确与否设计相应的奖励函数：
\begin{equation}
\begin{aligned}
R_{Judgment}(o_{ans}, o_{gt}) = \begin{cases}
1, & \text{if } o_{ans}=o_{gt} \\
0, & otherwise
\end{cases}
\end{aligned}
\end{equation}

Where $o_{ans}$ represents the binary response of the model and $o_{gt}$ is the correct answer. This simple reward design allows LLMs to understand recommendation data during reinforcement learning training, establishing associations for more complex rating prediction and explanation tasks in the following stages.
% 其中$o_{ans}$代表模型的二元响应，$o_{gt}$为正确答案。这一简单的奖励设计可以让大模型在强化学习训练中理解推荐数据，为后续更复杂的评级预测和解释生成推荐任务建立关联。

\subsubsection{Stage2: Choose for Rating}
We design a multi choice task to enable the LLM to learn the ability to predict ratings. The task instruction is "choose the most appropriate score from the following options," and we require the model's response to be one of the provided options. We have designed corresponding reward functions based on the ratings corresponding to the options provided by the model feedback and the ground truth ratings:
% 在该阶段，我们通过多元选择的任务设计让大模型学习评分预测的能力，任务指令为"choose the most appropriate score from the following options"，我们要求模型的响应必须是已提供的选项中的一个。详细提示词信息见附录A.1。我们根据模型反馈的选项所对应的分数与真实评分设计了相应的奖励函数：
\begin{equation}
\begin{aligned}
R_{Choose}(r_{ans}, r_{gt}) = 1-\frac{|r_{ans}-r_{gt}|}{4}
\end{aligned}
\end{equation}

Among them, $r_{ans}$ represents the rating corresponding to the options fed back by the model, and $r_{gt}$ represents the ground truth rating. During the training process, the closer the predicted rating of the model is to the ground truth rating, the more rewards it will receive. This reward mechanism helps to improve the model's ability to predict ratings.
% 其中$r_{ans}$代表模型反馈的选项所对应的评分，$r_{gt}$代表真实评分。在训练过程中，模型预测的评分越靠近真实评分，获得的奖励越多，这种奖励机制有助于提升模型评分预测的能力。

\subsubsection{Stage3: Generation for Explanation}
The first two stages are both discriminative tasks, and training through reinforcement learning can effectively enhance the model's reasoning ability. For the open response task of generating explanations, we not only need to consider how to generate smooth and reasonable explanations, but also need to consider coherence with the ratings. Too many optimization goals may lead to conflicts between rewards and fail to achieve good training results. To this end, we first use the DPO algorithm to train based on pre-organized preferences, with the aim of enabling the LLM to understand how to generate high-quality explanations. The following is an introduction to the collection method of preference pairs.
% 前两个阶段都属于判别式任务，通过强化学习训练可以较好的增强模型的推理能力。对于解释生成这一开放问答任务，我们不仅要考虑如何生成流畅合理的解释，还要考虑与评分的一致性。优化目标过多可能会导致奖励之间相互冲突，起不到良好的训练效果。为此，我们先采用DPO算法根据事先整理的偏好对进行训练，目的在于让大模型懂得如何生成高质量的解释。下面是关于偏好对采集方法的介绍。

Firstly, we use user and item profile information, as well as review data, as inputs for explainable recommendation tasks, and utilize the response of the GPT-3.5-turbo model as the chosen response. Using the LLM RL-2 model trained in the previous stage as a reference model, a certain proportion of random masks are applied to the effective information in the original input and passed to LLM RL-2. If the semantic similarity between the LLM output and the chosen response is below a certain threshold, we consider it a rejected response. 
% 首先，我们使用用户和商品配置信息以及评论数据作为可解释推荐任务的输入，借助GPT-3.5-turbo模型的响应作为chosen response。使用上一阶段训练得到的模型LLM RL-2作为参考模型，对原始输入中的有效信息进行一定比例的随机掩码传递给LLM RL-2，得到的输出如果和chosen response的语义相似度低于特定阈值，我们将其视为rejected response，数据收集过程可见算法1.

% \begin{algorithm}[H]
% \caption{Collecting Explanation Preference Pair for DPO Training}
% \begin{algorithmic}[1]  
%     \State \textbf{Required:} 
%     \State (1) An advanced model $LLM_{advanced}$
%     \State (2) User Profile $\{u_k\}$ and Item Profile $\{i_k\}$ with Review $\{r_k\}$
%     \State (3) Generating explanation prompt $P$
%     \State (4) Random Masker based on pos tagging $M$
%     \State \textbf{Input:} A LLM after Stage2 Training $LLM_{RL-2}$, Similarity threshold $\tau$, Mask ratio $\beta$
%     \State Initialize $D \leftarrow \emptyset$ //Init Preference Dataset
%     \State $K \leftarrow |\{u_k\}|$
%     \For{$k=1$ to $K$}
%         \State $x_k \leftarrow (u_k,i_k,r_k)$, $y_w \leftarrow LLM_{advanced}(P,x_k)$
%         \While{True}
%             \State $y \leftarrow LLM_{RL-2}(P,M(x_k,\beta))$
%             \If{$Sim(y,y_w) < \tau$}
%                 \State $y_l \leftarrow y$, $ D \leftarrow D \cup \{(x_k, y_w, y_l)\}$ 
%                 \State \textbf{break}  
%             \EndIf
%         \EndWhile
%     \EndFor
%     \State \textbf{Return} $D$ 
% \end{algorithmic}
% \label{algorithm}
% \end{algorithm}

After DPO training of LLM, we use the GRPO algorithm to train LLM to further improve the quality of explanation and the coherence of rating and explanation. The task instruction is "explain why the user would enjoy this item," and the model is required to output a simple and fluent response. At this stage, we designed length rewards to encourage the LLM to generate as concise responses as possible, accuracy rewards to drive the explanations generated by the LLM to align with standard answers, and coherence rewards to make the generated explanations and ratings as consistent as possible. The following is a specific introduction to the reward function.
% 对LLM进行DPO训练后，我们使用GRPO算法训练大模型以进一步提升解释的质量文本和评分一致性。任务指令为"explain why the user would enjoy this item", 并要求模型输出简单流畅的回复，提示词的详细信息见附录A.1。在该阶段，我们设计了长度奖励来鼓励大模型生成尽可能精简的回复，设计了准确性奖励来驱使大模型生成的解释向标准答案对齐，还设计了一致性奖励使生成的解释与评分尽可能一致。下面是关于奖励函数的具体介绍。

(1) Length reward: For length rewards, we classify them based on the length of the answers provided by LLM.
% (1) 长度奖励：对于长度奖励，我们根据LLM反馈的答案长度进行分级奖励。
\begin{equation}
\begin{aligned}
R_{Length}(t_{ans}) = \begin{cases}
1 & \text{if } len(t_{ans})<50 \\
0.5 & \text{if } 50 \leq len(t_{ans}) < 100 \\
0 & otherwise
\end{cases}
\end{aligned}
\end{equation}

Where $t_{ans}$ represents the textual representation of the generated answer.
% 其中$t_{ans}$表示生成答案的文本表示

(2) Accuracy reward: We use the semantic similarity between the answers fed back by LLM and the standard answers as an accuracy reward.
% 2. 准确性奖励：我们将大模型反馈的答案和标准答案的语义相似度作为准确性奖励
\begin{equation}
\begin{aligned}
R_{Accuracy}(t_{ans}, t_{gt}) = Similarity(T(t_{ans}), T(t_{gt}))
\end{aligned}
\end{equation}

Where $t_{ans}$ represents the text representation of the generated answer, $t_{gt}$ represents the text representation of the standard answer, and $T$ is the text encoder.
% 其中$t_{ans}$表示生成答案的文本表示，$t_{gt}$表示标准答案的文本表示，$T$为文本编码器

(3) Coherence reward: We use a BERT-based pre-trained sentiment classification model that takes the feedback from LLM as input and outputs an explanation based rating. Then evaluate the rating against the ground truth rating as a coherence reward.
% 3. 一致性奖励: 我们使用一个基于bert的预训练情感分类模型，将LLM反馈的答案作为输入，输出为基于解释的评分。然后将该评分与真实评分进行评估作为一致性奖励
\begin{equation}
\begin{aligned}
R_{Coherence}(t_{ans}, r_{gt}) = 1-\frac{|C(t_{ans})-r_{gt}|}{4}
\end{aligned}
\end{equation}

Where $t_ {ans}$ is the textual representation of the generated answer, $r_ {gt}$ is the ground truth rating. $C(\cdot)$ is the sentiment classification model with text as input and sentiment classification labels as output. Here, we use the rating as the label.
% 其中$t_{ans}$为生成答案的文本表示, $r_{gt}$为真实评分，$C(\cdot)$为一个情感分类模型，输入为文本，输出为情感分类标签，在这里我们将评分作为标签。

\subsection{Evaluation-Rating Coherence Evaluation}
The previous explanation-rating coherence evaluation work only considered the consistency between the generated explanation and the predicted rating but ignored the situation where both were distorted at the same time. In order to make the model more reliable, we also considered the accuracy with the ground truth rating during the evaluation and balanced between coherence and accuracy.
% 之前的解释-评分一致性评估工作只考虑了将生成解释与预测评分之间的一致性，却忽视了二者同时失真的情况，为了让模型更可信，我们在评估时还考虑了与真实评分的准确性，并在一致性和准确性之间进行了权衡。

Firstly, user and item profile information, as well as review data, are used as inputs for the model to be evaluated. Task instructions from stage-2 and stage-3 are used to predict ratings and generate explanations, respectively. In order to obtain explanation based ratings, we adopted three methods: one is to rate explanations based on GPT-3.5-turbo through instruction prompts, one is to rate explanations using a sentiment classification model, and the other is manual rating. We will denote the ground truth rating as $r_0$, the model-predicted rating as $r_1$, and the explanation based rating as $r_2$. The following is a coherence evaluation plan:
% 首先将用户和商品配置信息以及评论数据作为待评估模型的输入，分别使用阶段2和阶段3的任务指令来预测评分和生成解释，为了获得基于解释的评分，我们采用了三种方式，一种是基于GPT-3.5-turbo通过指令提示的方式对解释进行评分，一种是在第三阶段设计一致性奖励时使用的情感分类模型对解释进行评分，另一种是人工评分。我们将真实评分记作$r_0$,模型预测评分记作$r_1$,基于解释的评分记作$r_2$，下面给出一致性评估方案：

\begin{equation}
\begin{aligned}
&Acc = \frac{Accuracy + Coherence}{2} \\
&Accuracy = \frac{1}{2}*[(1-\frac{|r_0-r_1|}{4})+(1-\frac{|r_0-r_2|}{4})] \\
&Coherence = 1-\frac{|r_1-r_2|}{4}
\end{aligned}
\end{equation}

% Where $Acc$ is the final evaluation metric, and $\alpha$ is the balance weight between accuracy and coherence.
% 其中Acc为最终评估指标，$\alpha$为准确性和一致性的平衡权重

%% file: section/experiment.tex
\section{Experiment}
\subsection{Experiment Setting}

\begin{table}[htbp]
\centering
\caption{Dataset Statistics}
\label{table1}
\begin{tabular}{lrrrr}
\toprule
\textbf{Dataset} & \textbf{\# Users} & \textbf{\# Items} & \textbf{\# Train ($u$-$i$)} & \textbf{\# Test ($u$-$i$)} \\
\midrule
Baby & 19445 & 7050 & 3000 & 1000 \\
Sports & 35598 & 18357 & 3000 & 1000 \\
Clothing & 39387 & 23033 & 3000 & 1000 \\
\bottomrule
\end{tabular}
\end{table}

\subsubsection{Dataset} 
To evaluate the effectiveness of our method, we conducted experiments on three open-source datasets extracted from the Baby, Clothing\_Shoes\_and\_Jewelry, and Sports\_and\_outdoors subsets of the Amazon Review Dataset\footnote{https://jmcauley.ucsd.edu/data/amazon/links.html}. In addition, we used GPT-3.5-Turbo to generate user and item profile information as well as standard explanations based on metadata and review information. Considering that we train the model through curriculum learning, we collect different data from the source dataset in each stage, but with the same splitting method, we randomly divide each dataset into training, validation, and testing sets in a 3:1:1 ratio. Each record in the dataset contains a user ID, an item ID, a 1$\sim$5 rating, a task instruction, and a correct answer. The statistical information of the dataset is shown in Table \ref{table1}.
% 为了评估我们方法的有效性，我们使用了三个开源数据集上开展了实验，这三个数据集分别提取自Amazon Review Dataset中的Baby、Clothing_Shoes_and_Jewelry和Sports_and_Outdoors子集，此外，我们使用GPT-3.5-Turbo来根据元数据和评论信息生成用户和项目的配置信息以及标准解释。考虑到我们采用课程学习的方式训练模型，每个阶段从源数据集中使用不一样的数据，但切分方式相同，将每个数据集以3:1:1的方式随机分为训练集、验证集和测试集。数据集的每条记录包含一个用户ID、商品ID、一个1~5的评分、包含用户和商品配置信息以及评论数据的指令和正确答案。数据集的统计信息如表2所示

\subsubsection{Details of Baselines}
We compare our model’s performance against the following baselines:
% 我们将本模型的性能与以下基线模型进行了比较：
\begin{itemize}[leftmargin=*, noitemsep] 
\item \textbf{Att2Seq~\cite{dong2017learning}}: Utilizes an attention-based attribute-to-sequence model to generate reviews based on attribute information.
\item \textbf{NRT~\cite{li2017neural}}: Predicts ratings and generates abstractive tips for recommendations using multi-task learning to optimize parameters.
\item \textbf{PETER~\cite{li2021personalized}}: PETER is a personalized transformer model for explainable recommendation. It maps user and item IDs to the generated explanation text, connecting the IDs and words.
% \item \textbf{PEPLER~\cite{li2023personalized}} : PEPLER leverages pretrained transformer to generate explainable recommendations based on prompts that incorporate user and item ID vectors. There are several variants of PEPLER, in our experiment, we chose to use the continuous prompt learning version.
\item \textbf{CER~\cite{raczynski2023problem}}: Estimates the discrepancy between predicted ratings and explanation-based ratings to enhance rating-explanation coherency.
\item \textbf{XRec~\cite{ma2024xrec}}: Utilizes the encoded user/item embeddings from GNNs as implicit collaborative signals, which are then integrated into each layer of LLMs, enabling the generation of explanations.
\end{itemize}

Due to the fact that the dataset in our experiment has not been used in the baselines, here is an additional explanation of the collection method for the baseline model dataset. For Att2Seq, NRT, PETER, and CER models, we used the Sentires-Guide\footnote{https://github.com/lileipisces/Sentires-Guide} tool to extract sentiment quadruples from user reviews in the original dataset and trained and tested them using the same dataset segmentation method as the original paper. For the XRec model, we used the same training and testing samples as our own model, but the model also involves the injection of collaborative information, so we rigorously reproduced the paper. Construct an interaction graph based on the interaction information of the original dataset, and use trained GNNs to encode each user and item.
% 由于我们使用的数据集在基线中并没有被使用过，这里补充说明下基线模型数据集的收集方式。对于Att2Seq、NRT、PETER、PERLER和CER模型，我们使用Sentires-Guide工具从原始数据集的用户评论信息中抽取情感四元组，并按与原论文相同的数据集切分方式进行训练和测试分析。对于XRec模型，我们使用与本模型相同的训练和测试样本，但是该模型还涉及协同信息的注入，因此我们对论文进行了严格的复现。根据原始数据集的交互信息构建交互图，利用已经训练过的GNN为每个用户和商品进行编码。

\subsubsection{Evaluation Metrics}
In order to evaluate the explanation performance of the model, we followed XRec and adopted a set of metrics aimed at capturing the semantic explainability and stability of generated explanations. Traditional metrics based on n-gram syntax, such as BLUE~\cite{papineni2002bleu} and ROUGE~\cite{lin2004rouge}, are insufficient to achieve their goals due to their inability to fully capture semantic meanings. Specifically, we use GPTscore, BERTscore, BARTscore, BLEURT, and USR to measure explainability. Among them, GPTScore maintains consistency with human judgment by comparing the semantic similarity between generated explanations and ground truth explanations; BERTScore utilizes BERT's context embedding to calculate token-level cosine similarity; BARTScore utilizes the BART model to conceptualize evaluation as a text generation task, assigning scores based on the probability of regenerating reference text; BLEURT adopts a novel synthetic data pre-training method to enhance generalization; and USR measures the uniqueness of generated explanations by calculating the ratio of unique sentences to total sentences. To further evaluate the quality stability, we analyzed the standard deviation of these scores, where lower values indicate more consistent performance. The relevant evaluation results are shown in Table \ref{table3}.
% 为了评估模型的解释性能，我们遵循XRec工作采用了一套指标，旨在捕获生成解释的语义可解释性和稳定性。传统的基于n元语法的指标，如BLUE和ROUGE，由于无法完全捕获语义含义，因此不足以达到目的。具体来说，我们采用GPTscore、BERTscore、BARTscore、BLEURT和USR来衡量可解释性。其中，GPTScore通过比较生成的解释和真实解释之间的语义相似性，与人类判断保持一致; BERTScore利用 BERT的上下文嵌入来计算令牌级余弦相似性；BARTScore利用BART模型，将评估概念化为文本生成任务，根据重新生成参考文本的概率分配分数；BLEURT采用一种新颖的合成数据预训练方法来增强泛化；USR通过计算独特句子与总句子的比率来衡量生成解释的唯一性。为了进一步评估质量稳定性，我们分析了这些分数的标准差，值越低表示性能越一致。相关评估结果如表3所示。

To evaluate the performance of rating predictions, we used two commonly used metrics, Root Mean Square Error (RMSE) and Mean Absolute Error (MAE), to measure the deviation between predicted ratings and actual ratings. The performance evaluation of rating prediction is shown in Table \ref{table4}.
% 为了评估评级预测的性能，我们使用了两个常用的指标：均方根误差（RMSE）和平均绝对误差（MAE）来衡量预测评级和真实评级之间的偏差。关于评级预测的性能评估如表4所示。

In addition, we also evaluated the coherence of the model in terms of explanation and rating. The definition of the indicators is given in Section 4.3. We used three methods: GPT, a sentiment classification model, and manual evaluation. The coherence results are shown in Table \ref{table5}.
% 此外，我们还评估了模型在解释和评级方面的一致性，指标的定义在4.3节中已给出，我们采用了gpt、情感分类模型和手动评估三种方法，一致性结果如表5所示。

\subsubsection{Implementation Details}
All experiments were conducted on 8 NVIDIA RTX-4090 GPUs. We used the Qwen2.5-3B-Instruct model for the experiments, and the sentiment classification model used in the evaluation was the bert-base-multilingual-truncated-sentiment on Huggingface. The text encoding model used for the accuracy reward in Stage 3 was m3e-base. In GRPO, the sample number of responses for each prompt is set to 4, and the maximum prompt length is set to 512; the KL penalty coefficient $\beta$ in DPO is set to 0.1. The number of training samples for each stage is 3000, and the training time for each stage is approximately 12 hours; the test sample is 1000, and the inference time is about half an hour.
% 所有的实验均在8卡4090服务器上进行，我们使用Qwen2.5-3B-Instruct模型开展的实验，评估中使用的情感分类模型为huggingface上的bert-base-multilingual-uncased-sentiment，阶段三的准确性奖励所使用的文本编码模型为m3e-base。GRPO算法中每个prompt的回复采样数量设为4，最大输入长度设为512；DPO算法中KL惩罚系数 beta设为0.1
% 每个阶段的训练样本数为3000条，训练时长大概在12个小时左右；测试样本为1000条，推理时间在半小时左右。

\begin{table*}[ht]
\centering
\setlength{\tabcolsep}{3pt}
\renewcommand{\arraystretch}{1.1}
\caption{Explainability and Stability results on Three Datasets. "↑" means higher is better; "↓" means lower is better. Superscripts “P,” “R,” and “F1” denote Precision, Recall, and F1-Score, respectively. The subscript “std” indicates the standard deviation of each metric. Bold indicates the best results, while underlined denotes the second-best. (The significant improvements with p-value < 0.05 based on paired t-tests)}
\label{table3}
\resizebox{\textwidth}{!}{  % 自动按页面宽度缩放
\begin{tabular}{l|l|ccccccc|cccccc}
\toprule
\multirow{2}{*}{Category} & \multirow{2}{*}{Models} & \multicolumn{7}{c|}{\textbf{Explainability} $\uparrow$} & \multicolumn{6}{c}{\textbf{Stability} $\downarrow$} \\
 & & GPT$_\text{score}$ & BERT$^P_\text{score}$ & BERT$^R_\text{score}$ & BERT$^{F1}_\text{score}$ & BART$_\text{score}$ & BLEURT & USR & GPT$_\text{std}$ & BERT$^P_\text{std}$ & BERT$^R_\text{std}$ & BERT$^{F1}_\text{std}$ & BART$_\text{std}$ & BLEURT$_\text{std}$ \\
\midrule
\multicolumn{15}{c}{\textbf{Baby}} \\
\midrule
\multirow{5}{*}{Baseline} 
& NRT          & 22.36 & 0.1515 & 0.0998 & 0.1260 & -5.2312 & -1.0946 & 0.0094 & 17.09 & 0.1525 & 0.1506 & 0.1351 & 1.0305 & 0.2325 \\
& Att2Seq      & 26.19 & 0.1356 & 0.0887 & 0.1119 & -5.2568 & -1.0701 & 0.0437 & 21.14 & 0.1827 & 0.1608 & 0.1465 & 1.0432 & 0.2667  \\
& PETER        & 45.11 & 0.3278 & 0.2305 & 0.2784 & -4.7111 & -0.8737 & 0.5338 & 26.25 & 0.2051 & 0.2103 & 0.1889 & 1.0288 & 0.4010 \\
% PEPLER       & 24.95 & 0.0951 & 0.0134 & 0.0530 & \underline{-3.9049} & -1.2336 & 0.6782 & 21.49 & 0.1731 & 0.2077 & 0.1543 & 1.0600 & \underline{0.1476} \\
& CER        & 46.11 & 0.3154 & 0.2292 & 0.2716 & -4.7099 & -0.8669 & 0.6078 & 26.02 & 0.2045 & 0.2076 & 0.1863 & 1.0304 & 0.4408 \\
& XRec-3B    & 82.38 & \underline{0.4273} & \textbf{0.4422} & \textbf{0.4354} & -3.9362 & -0.0987 & 0.9989 & 9.16 & 0.1171 & 0.1153 & 0.1111 & 0.6086 & 0.2012 \\
\midrule
& Qwen2.5-7B-Instruct & 84.89 & 0.2313 & 0.3559 & 0.2936 & \underline{-3.6146} & 0.1030 & \textbf{1.0000} & \underline{4.05} & \underline{0.1017} & \underline{0.0794} & \underline{0.0821} & \underline{0.4526} & 0.1298 \\
& Qwen2.5-14B-Instruct & \underline{85.10} & 0.2483 & \underline{0.3660} & 0.3074 & -3.6421 & \underline{0.1045} & \textbf{1.0000} & \textbf{3.92} & \textbf{0.0810} & \textbf{0.0779} & \textbf{0.0730} & \textbf{0.4523} & \underline{0.1281} \\
\midrule
\rowcolor{gray!10}
Ours & Curr-RLCER-3B  & \textbf{86.98} & \textbf{0.4533} & 0.3569 & \underline{0.4054} & \textbf{-3.5523} & \textbf{0.2377} & \textbf{1.0000} & 7.85 & 0.1027 & 0.0957 & 0.0934 & 0.5072 & \textbf{0.0800} \\
\midrule
\multicolumn{15}{c}{\textbf{Sports}} \\
\midrule
\multirow{5}{*}{Baseline} 
& NRT          & 26.62 & 0.0319 & 0.0574 & 0.0410 & -5.6958 & -1.1106 & 0.0595 & 21.12 & 0.3362 & 0.1572 & 0.2165 & 1.0707 & 0.2767 \\
& Att2Seq      & 26.18 & -0.0326 & 0.0489 & 0.0033 & -5.7100 & -1.1307 & 0.0383 & 21.58 & 0.3653 & 0.1521 & 0.2295 & 1.0720 & 0.2312  \\
& PETER        & 43.54 & 0.2579 & 0.1778 & 0.2170 & -5.1567 & -0.8970 & 0.4670 & 25.85 & 0.2159 & 0.1948 & 0.1822 & 1.0723 & 0.3844 \\
% PEPLER       & 28.83 & 0.0828 & 0.0542 & 0.0674 & \textbf{-3.4767} & -1.2145 & 0.7108 & 24.03 & 0.1353 & 0.2129 & 0.1364 & 1.2216 & \underline{0.1620} \\
& CER        & 46.15 & 0.2709 & 0.1779 & 0.2236 & -5.1481 & -0.8999 & 0.5122 & 25.76 & 0.2037 & 0.1963 & 0.1778 & 1.0716 & 0.3759 \\
& XRec-3B    & 79.12 & \underline{0.4315} & \textbf{0.4353} & \textbf{0.4341} & -3.9651 & -0.1247 & \textbf{1.0000} & 10.21 & 0.1154 & 0.1161 & 0.1108 & 0.6184 & 0.1778 \\
\midrule
\multirow{2}{*}{Base LLM}
& Qwen2.5-7B-Instruct & 84.43 & 0.2373 & 0.3512 & 0.2944 & \underline{-3.7328} & 0.1008 & \textbf{1.0000} & \underline{4.21} & 0.1044 & \underline{0.0784} & \underline{0.0836} & \textbf{0.4268} & 0.1310 \\
& Qwen2.5-14B-Instruct & \underline{84.54} & 0.2559 & \underline{0.3609} & 0.3087 & -3.7483 & \underline{0.1016} & \textbf{1.0000} & \textbf{4.04} & \underline{0.1009} & \textbf{0.0750} & \textbf{0.0806} & \underline{0.4271} & \underline{0.1301} \\
\midrule
\rowcolor{gray!10}
Ours & Curr-RLCER-3B  & \textbf{86.37} & \textbf{0.4739} & 0.3537 & \underline{0.4138} & \textbf{-3.6804} & \textbf{0.2387} & \textbf{1.0000} & 7.43 & \textbf{0.0983} & 0.0894 & 0.0873 & 0.4855 & \textbf{0.1015} \\
\midrule
\multicolumn{15}{c}{\textbf{Clothing}} \\
\midrule
\multirow{5}{*}{Baseline} 
& NRT          & 31.09 & 0.1748 & 0.1009 & 0.1381 & -5.4719 & -1.0274 & 0.0026 & 23.78 & 0.1302 & 0.1593 & 0.1289 & 1.0462 & 0.3112 \\
& Att2Seq      & 35.10 & 0.2079 & 0.1209 & 0.1643 & -5.3831 & -1.0108 & 0.0303 & 25.76 & 0.1431 & 0.1683 & 0.1379 & 1.0539 & 0.3145  \\
& PETER        & 50.83 & 0.3489 & 0.2586 & 0.3029 & -4.8712 & -0.7521 & 0.2737 & 26.19 & 0.2019 & 0.2267 & 0.1945 & 1.0913 & 0.5012 \\
% PEPLER       & 35.96 & 0.1332 & -0.0029 & 0.0625 & \underline{-3.7008} & -1.2115 & 0.4579 & 27.93 & 0.1663 & 0.2425 & 0.1648 & 0.9730 & \underline{0.1435} \\
& CER        & 52.89 & 0.3687 & 0.2592 & 0.3129 & -4.8688 & -0.7577 & 0.1790 & 27.50 & 0.2103 & 0.2349 & 0.2031 & 1.1084 & 0.5104 \\
& XRec-3B    & 81.93 & \underline{0.4043} & \textbf{0.4159} & \textbf{0.4108} & -3.9113 & -0.1476 & \textbf{1.0000} & 11.72 & 0.1246 & 0.1201 & 0.1166 & 0.5979 & 0.2180 \\
\midrule
\multirow{2}{*}{Base LLM}
& Qwen2.5-7B-Instruct & 84.07 & 0.2228 & 0.3285 & 0.2759 & -3.7088 & 0.0616 & \textbf{1.0000} & \textbf{3.74} & \textbf{0.0948} & \textbf{0.0790} & \textbf{0.0771} & \underline{0.4057} & \underline{0.1483} \\
& Qwen2.5-14B-Instruct & \underline{84.29} & 0.2424 & \underline{0.3432} & 0.2931 & \underline{-3.7026} & \underline{0.0620} & \textbf{1.0000} & \underline{4.79} & \underline{0.1015} & \underline{0.0814} & \underline{0.0849} & \textbf{0.4033} & 0.1505 \\
\midrule
\rowcolor{gray!10}
Ours & Curr-RLCER-3B  & \textbf{86.18} & \textbf{0.4488} & 0.3243 & \underline{0.3866} & \textbf{-3.6707} & \textbf{0.2444} & \textbf{1.0000} & 8.19 & 0.1042 & 0.0940 & 0.0912 & 0.4351 & \textbf{0.0811} \\
\bottomrule
\end{tabular}
}
\end{table*}

\subsection{Model Performance}
\subsubsection{Evaluation of Explanation}
To demonstrate the superiority of our model in explainability and stability, we conducted a comparative analysis of five baseline methods in three datasets. The results are summarized in Table \ref{table3}. Based on observations, it can be observed that Curr-RLCER exhibits excellent performance in both explainability and stability. Compared with the best baseline model XRec, although our method has a slightly lower recall rate when validated using the Bert model, this may be due to the introduction of additional collaborative information in XRec, which covers more features and provides a deeper understanding of recommendation behavior compared to using only text modality. In other indicators, our model performs better and has stronger stability. This demonstrates the effectiveness of the Stage-3 training in our method, where the model understands how to generate high-quality explanations through training with DPO and GPRO. In addition, considering that most baseline models use traditional deep learning architectures, in order to more intuitively demonstrate the effectiveness of our approach, we conducted comparative tests with backbone models of different parameter sizes in the same series. The results showed that Curr-RLCER can effectively improve the model's explainability, but its stability may deteriorate.
% 为了证明我们的模型在可解释性和稳定性方面的优越性，我们在三个数据集中对五种基线方法进行了比较分析。结果总结在表 3 中。根据观察可以发现，Curr-RLCER在可解释性和稳定性方面都表现出卓越的性能。与最优秀的基线模型XRec相比，虽然我们方法在使用Bert模型验证时的召回率略低于XRec,这可能是因为XRec中还引入了额外的协同信息，从而覆盖更多的特征，对推荐行为的理解相比于只使用文本模态更加深入。在其他指标中我们的模型均表现的更加优异，并且稳定性更强。这证明了我们方法中第三个阶段训练的有效性，通过dpo和gpro的训练，模型理解如何生成高质量的解释。
% 此外，考虑到基线模型大多采用传统的深度学习架构，为了更直观的证明本方案的有效性，我们与同系列不同参数大小的骨干模型进行了对比测试，结果表明，Curr-RLCER可以有效的提升模型的解释能力，不过稳定性会变差。

\begin{table*}[ht]
\centering
\caption{Rating Prediction Performance Comparison Across Datasets. “R” means RMSE, and “M” means MAE.}
\label{table4}
\resizebox{\textwidth}{!}{  % 自动按页面宽度缩放
\begin{tabular}{l|cc|cc|cc}
\toprule
\multirow{2}{*}{Models} & \multicolumn{2}{c|}{\textbf{Baby}} & \multicolumn{2}{c|}{\textbf{Sports}} & \multicolumn{2}{c}{\textbf{Clothing}} \\
 & \textbf{R$\downarrow$} & \textbf{M$\downarrow$} & \textbf{R$\downarrow$} & \textbf{M$\downarrow$} & \textbf{R$\downarrow$} & \textbf{M$\downarrow$} \\
\midrule
NRT & 1.1164 & 0.8018 & 1.0680 & 0.7308 & 1.0528 & 0.7828 \\
PETER & 1.0625 & 0.7982 & 0.9490 & 0.7217 & 1.0615 & 0.8218 \\
CER & 1.0613 & 0.8343 & 0.9518 & 0.7382 & 1.0612 & 0.8349 \\
Qwen2.5-7B-Instruct & 0.8426 & 0.5480 & 1.0154 & 0.7010 & 0.9143 & 0.5840 \\
Qwen2.5-14B-Instruct & 0.8277 & 0.5197 & 0.8915 & 0.6660 & 0.8882 & 0.5818 \\
\rowcolor{gray!10}
Curr-RLCER & \textbf{0.6899} & \textbf{0.33} & \textbf{0.6943} & \textbf{0.3340} & \textbf{0.6099} & \textbf{0.3080} \\
\bottomrule
\end{tabular}
}
\end{table*}

\subsubsection{Evaluation of Rating Prediction}
To demonstrate the accuracy of our model in rating prediction, we conducted a comparative analysis of all baseline methods with rating prediction capability in three datasets. The results are shown in Table \ref{table4}. Based on observations, it can be observed that our method has excellent rating prediction ability, with RMSE improved by 34.9\%, 27.1\%, and 42.5\% compared to the best baseline model on three datasets; MAE improved by 58.6\%, 53.7\%, and 60.6\% compared to the best baseline model. And compared to the backbone models with higher parameters in the same series, the rating prediction ability has also been significantly improved. This proves the effectiveness of Stage 1 and Stage 2 of training in our method. Through CTR prediction and rating prediction, the model can understand the sentiment in reviews based on profile information and make accurate rating predictions.
% 为了证明我们的模型在评级预测方面的准确性，我们在三个数据集中对所有基线方法中含有评级预测能力的模型进行了比较分析。结果如表4所示。根据观察可以发现，我们的方法具有非常良好的评级预测能力，在三个数据集上RMSE指标相比于最好的基线模型提升了34.9%，27.1%，42.5%；MAE指标相比于最好的基线模型提升了58.6%，53.7%，60.6%。并且相比于同系列更高参数量的骨干模型，评分预测能力亦有较为明显的提升。这证明了我们方法中前两个阶段训练的有效性，通过CTR预测和评级预测，模型可以根据配置信息理解评论信息中的情感，并做出准确的评分预测。

\subsubsection{Evaluation of Coherence}
To demonstrate that our model can improve the coherence between explanation and rating, we conducted consistency evaluations on three datasets using the custom metrics introduced in Section 3.3 and compared them with NRT, PETER, and CER models that include both explanation generation and rating prediction tasks. The results are shown in Table \ref{table5}. Based on observations, it can be observed that our method can significantly improve the consistency between explanatory ratings and predictive ratings while maintaining a high degree of rating accuracy. This demonstrates the effectiveness of the coherence reward mechanism in stage-3 GRPO training, enabling the model to better match user ratings when generating explanations.
% 为了证明我们的模型能够提升解释-评分的一致性，我们在三个数据集上使用4.3节中介绍的自定义指标进行了一致性评估，并与同时包含解释生成和评级预测任务的NRT、PETER、CER模型进行了对比分析。结果如表5所示。根据观察可以发现，我们的方法可以大幅提高解释评级和预测评级之间的一致性，并且保持高度的评级准确性。这证明了阶段三GRPO训练中一致性奖励机制的有效性，使得模型在生成解释时能够更加符合用户评分。

Here is the detailed description of our manual annotation method. There were a total of 5 researchers involved in manual annotation. We extracted 200 test samples from each dataset and utilized Curr-RLCER and other baseline models to generate explanations and predict ratings. The researchers will rate each explanation generated by each model ranging from 1 to 5, based on sentiment expression. Afterwards, we will summarize the scoring results of each researcher as the explanation-based rating for the test samples. Finally, we will calculate the coherence indicator by combining the explanation-based rating with the ground truth rating and the predicted rating.
% 这里说明下我们人工标注的方法。参与人工达标的研究人员共5人，我们对每个数据集各抽出200条测试样例，然后分别使用Curr-RLCER和其他基线模型生成解释和预测评分。研究人员的工作是对各个模型产生的每一条解释根据情感表达从1~5之间进行打分。之后我们汇总每个研究人员的打分结果，作为测试样例的解释评分。将其与真实评分、预测评分一起计算一致性分数，作为人工评估的测试结果。

\begin{table*}[ht]
\centering
\caption{Coherence Performance Comparison Across Datasets. We use GPT-3.5-Turbo, a BERT-based sentiment classifier, and human annotations for explanation-rating coherence evaluation.}
\label{table5}
\resizebox{\linewidth}{!}{  % 自动按页面宽度缩放
\begin{tabular}{l|ccc|ccc|ccc}
\toprule
 & \multicolumn{3}{c|}{\textbf{GPT}} & \multicolumn{3}{c|}{\textbf{Bert-Classifier}} & \multicolumn{3}{c}{\textbf{Human Annotator}} \\
 & Baby & Sports & Clothing & Baby & Sports & Clothing & Baby & Sports & Clothing \\
\midrule
NRT & 0.6824 & 0.7124 & 0.7541 & 0.8049 & 0.7549 & 0.8217 & 0.7028 & 0.7386 & 0.7324 \\
PETER & 0.7013 & 0.7226 & 0.7191 & 0.8119 & 0.8006 & 0.7994 & 0.6915 & 0.6716 & 0.7053 \\
CER & 0.6988 & 0.6714 & 0.7025 & 0.7996 & 0.8081 & 0.7973 & 0.7561 & 0.7084 & 0.7261 \\
\rowcolor{gray!10}
Curr-RLCER & \textbf{0.8334} & \textbf{0.8516} & \textbf{0.8679} & \textbf{0.9191} & \textbf{0.9036} & \textbf{0.9187} & \textbf{0.8617} & \textbf{0.8932} & \textbf{0.9008} \\
\bottomrule
\end{tabular}
}
\end{table*}

\subsection{Ablation Study}
In this section, we conducted ablation studies to explore the impact of two key components in the model: (1) the effect of DPO training on the model's explanatory power and (2) the effect of coherence reward mechanisms on the model's explanation rating prediction consistency. For this purpose, we compared three model variants: 1) a complete model with all features 2) \textbf{w/o DPO Training}: omitting preference pair collection and DPO training processes, and 3) \textbf{w/o Coherence Reward}: the reward model in stage-3 does not include coherence rewards. In order to make the ablation assessment more accurate, we conducted ablation experiments on the Baby and Clothing datasets. The experimental results are shown in Table \ref{table6}.
% 在本节中，我们进行了消融研究，以探索模型中两个关键组成部分的影响：（1）DPO训练对模型解释能力的影响 （2）一致性奖励机制对模型解释-评级预测一致性的影响。 为此，我们比较了三个模型变体：1）具有所有功能的完整模型 2）w/o DPO Training: 省略了偏好对收集和DPO训练过程 3）w/o Coherence Reward: 阶段三中的奖励模型中不包含一致性奖励。为了使消融评估更准确，我们在Baby和Clothing两个数据集上开展了消融实验。实验结果见表6.

\begin{table*}[ht]
\centering
\caption{Ablation Study Results. 1) w/o DPO Training: Omitting preference pair collection and DPO training processes. 2) w/o Coherence reward: The reward model in stage-3 does not include coherence rewards.}
\label{table6}
\resizebox{\linewidth}{!}{
\begin{tabular}{l|ccccc|ccccc}
\toprule
\multirow{2}{*}{Ablations} & \multicolumn{5}{c|}{\textbf{Baby}} & \multicolumn{5}{c}{\textbf{Clothing}} \\
 & MAE$\downarrow$ & RMSE$\downarrow$ & BERT$^{F1}_{score}$ $\uparrow$ & BERT$^{F1}_{std}$ $\downarrow$ & Acc$\uparrow$ & MAE$\downarrow$ & RMSE$\downarrow$ & BERT$^{F1}_{score}$ $\uparrow$ & BERT$^{F1}_{std}$ $\downarrow$ & Acc$\uparrow$ \\
\midrule
Curr-RLCER & 0.33 & 0.6899 & 0.4054 & \textbf{0.0934} & \textbf{0.9191} & \textbf{0.3080} & \textbf{0.6099} & 0.3866 & 0.0912 & \textbf{0.9187} \\
w/o Coherence Reward & \textbf{0.3140} & \textbf{0.6588} & \textbf{0.4224} & 0.0979 & 0.8307 & 0.3350 & 0.6656 & \textbf{0.4101} & 0.0921 & 0.8489 \\
w/o DPO Training & 0.3350 & 0.7036 & 0.3834 & 0.0985 & 0.8934 & 0.3270 & 0.6731 & 0.3542 & \textbf{0.0864} & 0.9072 \\
\bottomrule
\end{tabular}
}
\end{table*}

According to observations, it can be found that after removing the DPO training process, the model's explanation generation ability will significantly decrease, and its rating prediction ability will also be affected to some extent. This is likely due to the model's lack of understanding of how to generate high-quality explanations during open generation tasks in GRPO training, the model focuses on how to make the generated explanations more concise and accurate, and the coherence reward cannot play its role well. Therefore, in the process of improving the explanation generation ability, the disaster forgetting phenomenon occurs, weakening the rating prediction ability learned in the previous stage.
% 根据观察可以发现，删除了dpo训练过程后，模型的解释生成能力会大幅下降，并且评级预测能力也会受到一定影响，这应该是由于模型在缺乏如何生成高质量解释的理解的前提下进行开放式生成任务的gpro训练时，模型将训练中心关注到如何使生成的解释更加精简和准确，一致性奖励不能很好的发挥其作用，于是在提升解释生成能力的过程中出现了灾难遗忘现象，弱化了先前阶段学习到的评级预测能力。

In addition, after removing the coherence reward mechanism, the explanation generation ability of the model will be slightly improved, but the coherence is not as good as the complete model. This is because the model will only focus on learning how to generate high-quality explanations while ignoring the consistency between explanations and ratings.
% 此外，在删除一致性奖励机制之后，模型的解释生成能力会有小幅度的提升，但是一致性方面大不如完整模型。这是因为模型将只关注学习如何生成高质量的解释，却忽视了解释和评级之间的一致性，在评估指标上体现的非常明显。

\subsection{Model Robustness about Explainability}
\begin{figure}[t]
    \centering
    \includegraphics[width=\linewidth]{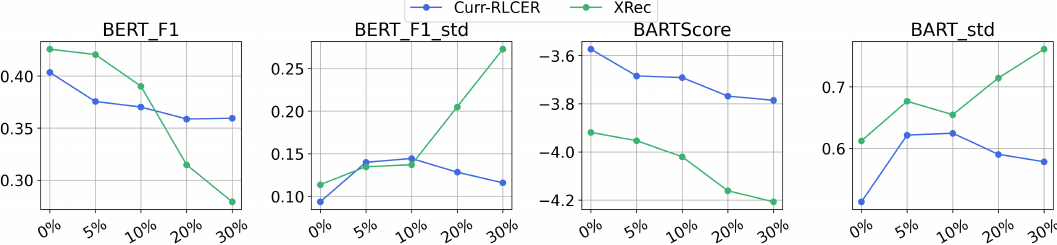}
    \caption{Robustness Experiment comparing Curr-RLCER with XRec in different noise ratios}   
    \label{fig4}
\end{figure}

In order to evaluate the generalization ability of the model in terms of explainability, we conducted a robustness comparison analysis between Curr-RLCER and XRec, which had the best explainability performance in all baselines. We add different proportions of noise to the input and demonstrate the robustness of the model based on the performance changes in generating explanations under different levels of noise. Specifically, we sampled 500 data points in the test set of the Baby dataset and added noise to the input instructions of these samples through random entity masking. The masking ratio was set to five groups: raw data (0\%), 5\%, 10\%, 20\%, and 30\%. Fig \ref{fig4} shows the changes in performance indicators of two models under different levels of noise. The evaluation results highlight several key findings:
% 为了评估模型在可解释性方面的泛化能力，我们对Curr-RLCER模型和基线中可解释性表现最好的XRec模型进行了鲁棒性对比分析。我们在输入中添加不同比例的噪声，根据在不同加噪程度下模型生成解释的性能变化来说明模型的鲁棒性。具体来说，我们在Baby数据集的测试集中采样了500条样本数据，并对这些样本的输入指令通过随机实体掩码的方式添加噪声，掩码比例一共设置了五组：原始数据（0%）、5%、10%、20%和30%。图4为在不同噪声下两个模型的性能指标变化情况。

\begin{itemize}[leftmargin=*, noitemsep] 
    \item As the noise ratio continues to increase, our model Curr-RLCER exhibits excellent noise resistance, and both semantic explainability and text generation stability do not undergo drastic changes with the increase of input noise. This trend reflects that our model has good robustness.
% 随着噪声比例的不断增加，我们的模型Curr-RLCER表现出了非常良好的抗噪能力，无论是语义上的可解释性还是文本生成的稳定性都未跟随输入噪声的增加而发生剧烈变化。这一趋势反映了我们模型具有较好的鲁棒性。

    \item In both noiseless and low-noise scenarios, our model performs slightly worse than XRec in terms of Bert\_F1 score, which may be due to the fact that XRec introduces collaborative signals by trained GNN, resulting in more effective information contained in the input. However, as the number of masks increased, XRec showed a significant decline in all evaluation indicators. The reason why our model can maintain robustness is that the reward design in reinforcement learning training can encourage the model to provide preference answers even in the presence of input disturbances.
% 在无噪声和低噪声场景下，我们模型在Bert_F1分数上略逊于XRec模型，这可能是因为XRec通过检索路径的方式引入了协同信号，输入所包含的有效信息量更多。但是随着掩码数量的增加，XRec在各项指标均体现出了明显的下降。而我们的模型能够保持稳健的原因是在强化学习训练的奖励设计可以鼓励模型在存在输入扰动时仍能给出偏好回答。
\end{itemize}

%% file: section/related.tex
\section{Related Work}
\subsection{Reinforcement Learning from Human Feedback}
Reinforcement Learning from Human Feedback (RLHF) has emerged as a crucial technique for incorporating human preference signals into machine learning methods and models~\cite{dong2024rlhf}. RLHF frameworks can be broadly categorized into deep RL-based approaches and direct preference learning approaches. In deep RL-based methods, a reward model is first constructed, after which Proximal Policy Optimization (PPO)~\cite{schulman2017proximal} is employed to optimize the reward signals with KL regularization~\cite{ouyang2022training}. While the direct preference learning approaches optimize a designed loss target on the offline preference dataset directly, eliminating the need for a separate reward model~\cite{tang2024generalized}. Recent extensions like Step-DPO~\cite{lai2024step} further decompose preference alignment into multi-step optimization. Besides, Dong et al.~\cite{dong2023raft} introduced rejection sampling fine-tuning to align LLMs with human preferences without requiring reinforcement learning pipelines. 

\subsection{Reasoning Model with LLM}
In recent years, with the continuous advancement of technologies such as Tree of Thoughts~\cite{yao2023tree} and Process-Supervised Learning~\cite{lu2024autopsv}, reasoning models have made significant progress. With OpenAI introducing RLM in their o1 models~\cite{wainwright2023instructgpt}, many works have explored various post-training combinations of reinforcement learning and supervised fine-tuning to further improve inference capabilities. Deepseek R1-Zero~\cite{guo2025deepseek} demonstrates that direct reinforcement learning of LLM still achieves well inference ability through the Group Relative Policy Optimization (GRPO) algorithm that does not rely on SFT. A large amount of work has shown the potential of using reinforcement learning to train small-scale RLMs in the fields of mathematics, logic, and coding ~\cite{wei2025swe,xie2025logic}. % In the recommendation domain, although some researchers have attempted to leverage LLM-based reasoning models to uncover user preferences and perform factual reasoning (e.g., Reason4Rec~\cite{fang2025reason4rec}, ReaRAG~\cite{huang2025rag}), these approaches face challenges of limited generalization capability and a lack of consistency. In this work, we combine the reasoning strengths of LLMs with curriculum reinforcement learning and introduce a dedicated explanation–rating coherence reward mechanism, enabling the model to improve explanation quality while better aligning rating prediction with explanation generation.
% 近年来，随着思维链、过程监督学习等技术的不断进步，推理模型取得了重大发展。随着OpenAI在其o1模型中引入了RLM，许多工作探索了强化学习和监督微调的各种训练后组合，以进一步提高推理能力。Deepseek R1-Zero通过不依赖于sft的组相对策略优化（GRPO）算法，表明直接对LLM进行强化学习依然获得良好的推理能力。目前已有大量工作表明在数学、逻辑和编码领域使用强化学习训练小规模RLM的潜力。
% 在推荐领域，虽有相关学者尝试基于LLM推理模型来挖掘用户偏好和事实推理【Reason4Rec、ReaRAG】，但面临泛化能力不足与一致性缺失的问题。本工作将LLM推理优势与课程强化学习结合，引入专门的解释-评分一致性奖励机制，使模型在提升解释质量的同时，更好地对齐评分预测与解释生成结果。

\subsection{Explainable Recommendation}
Explainable recommendation has the reasoning ability to explore user recommendation behavior logic, thereby enhancing the personalization and transparency of recommendation systems, which has significant research significance. In recent years, more and more research has focused on how to provide good explanations to improve system effectiveness and user satisfaction. Early methods mainly used predefined templates to generate explanations~\cite{li2021caesar} or extract logical reasoning rules from recommendation models~\cite{chen2021neural,shi2020neural}. With the continuous development of deep learning and the advancement of natural language generation technology, many works adopt attention mechanisms and recursive neural networks~\cite{li2017neural,dong2017learning} to generate natural language explanations. Currently, the most popular approach is to use LLMs for explainable recommendations~\cite{ma2024xrec,li2025g,qiu2024unveiling}. The good semantic understanding and generation ability of large language models can better improve the performance of model interpretation and generation. The existing explainable recommendation models mainly have the following two problems: (1) excessive reliance on high-quality supervised data and poor generalization and (2) only considering how to improve the quality of generated text, ignoring the coherence between predicted scores and explanations. This work uses reinforcement learning to train LLMs, thereby reducing reliance on supervised data; adds explanation rating coherence rewards during the training process; and conducts a comprehensive evaluation from both accuracy and coherence perspectives.
% 可解释推荐具有挖掘用户推荐行为逻辑的推理能力，从而提升推荐系统的个性化程度和透明度，具有非常重要的研究意义。近年来，越来越多的研究集中在如何提供良好的解释以提高系统有效性和用户满意度。早期方法主要使用预定义的模板生成解释或从推荐模型中提取逻辑推理规则。随着深度学习的不断发展和自然语言生成技术的进步，很多工作采用注意力机制和递归神经网络等技术，生成自然语言解释。目前，较为流行的是使用大模型进行可解释推荐。大语言模型良好的语义理解和生成能力可以更好地提高模型解释生成的性能。现有的可解释推荐模型主要存在以下两个问题：（1）过度依赖高质量监督数据，泛化性较差 （2）只考虑如何提高生成文本的质量，忽视了预测评分和解释的一致性。本工作使用强化学习来训练大模型，从而减轻对监督数据的依赖；在训练过程中添加解释-评分一致性奖励，并从准确性和一致性两个角度出发进行综合评估。

%% file: section/conclusion.tex
\section{Conclusion}
In this work, we introduce a reinforcement learning based LLM explainable recommendation framework, Curr-RLCER. While improving the model's explainable recommendation and rating prediction abilities, we also consider the issue of inconsistent explanations and ratings and design a coherence reward mechanism and coherence evaluation scheme. In terms of operational implementation, we adopt a curriculum learning training approach, conducting multi-task training from easy to difficult to alleviate the disaster forgetting problem of LLMs in multi-task learning; we adopt reinforcement learning training methods to alleviate the problem of traditional explainable recommendation schemes overly relying on high-quality supervised data. Extensive experimental results have shown that our method is effective.
% 在本工作中，我们介绍了一种基于强化学习的大模型可解释推荐框架Curr-RLCER，在提高模型可解释推荐和评分预测能力的同时，我们还考虑了解释和评分不一致的问题，并设计了一套一致性奖励机制和一致性评估方案。在操作实现上，我们采用课程学习的训练方式，从易到难的开展多任务训练，以缓解大模型在多任务学习中的灾难遗忘问题；并且采用强化学习的训练方式可以缓解传统可解释推荐方案过分依赖高质量监督数据的问题。广泛的实验结果表明，我们的方法无论是在解释生成还是评分预测上均优于以前最先进的方法，而且在解释和评分上表现出高度的一致性。此外，我们还进行了模型鲁棒性测试，实验结果表明，我们的模型在解释生成方面具有良好的抗噪能力。

\noindent\textbf{Acknowledgements.}
This work was supported in part by the National Natural Science Foundation of China under Grant No. 62276110, No. 62172039 and in part by the fund of Joint Laboratory of HUST and Pingan Property \& Casualty Research (HPL). The authors would also like to thank the anonymous reviewers for their comments on improving the quality of this paper.

\noindent\textbf{Disclosure of Interests.}
The authors have no competing interests to declare that are relevant to the content of this article.